# Pressure-induced decoupling of the pseudogap and superconductivity in a bilayer cuprate

Yanghao Meng[1, 2], Wenjin Mao[1, 2], Liucheng Chen[1, †], Elbert E. M. Chia[3], Yifeng Yang[1, 2], Tao Xiang[1, 2], Jianlin Luo[1, 2], Lin Zhao[1, 2, †], Xingjiang Zhou[1, 2], Xiaohui Yu[1, 2, 4, †], Xinbo Wang[1, 2, †]

[1] *Beijing National Laboratory for Condensed Matter Physics, Institute of Physics, Chinese Academy of Sciences, Beijing 100190, China*

[2] *School of Physical Sciences, University of Chinese Academy of Sciences, Beijing 100190, China*

[3] *Division of Physics and Applied Physics, School of Physical and Mathematical Sciences, Nanyang Technological University, Singapore 637371, Singapore*

[4] *Songshan Lake Materials Laboratory, Dongguan, Guangdong 523808, China*

[†]*Corresponding authors: chenliucheng@iphy.ac.cn, lzhao@iphy.ac.cn, yuxh@iphy.ac.cn, xinbowang@iphy.ac.cn*

**Abstract**

The physical origin of the pseudogap in cuprate superconductors, whether a precursor to superconductivity or a distinct intertwined order, remains an open question. Although hydrostatic pressure provides a clean tuning parameter, probing the relevant energy scales under compression has remained experimentally elusive. Here, we employ ultrafast optical spectroscopy to track the non-equilibrium quasiparticle dynamics of slightly underdoped $Bi_2Sr_2CaCu_2O_{8+\delta}$ up to 37 GPa. By disentangling the pseudogap and superconducting relaxation channels in the time domain, we uncover a contrasting pressure evolution: despite an increase in carrier density, the pseudogap thermal onset ($T^*$) rises monotonically, whereas its energy gap ($\Delta_{PG}$) is continuously suppressed. Conversely, the critical temperature ($T_c$) and the superconducting gap ($\Delta_{SC}$) exhibit a synchronized dome-like trajectory. The effective coupling ratio $2\Delta_{SC}/k_BT_c$ collapses abruptly from ~10 toward the weak-coupling $d$-wave limit, signifying a crossover from a phase-fluctuating two-dimensional regime to a globally coherent three-dimensional state. Upon compression to 37 GPa, the superconducting condensate is completely quenched into an insulating-like state, yet the pseudogap persists. The extended phase diagram reveals an unambiguous decoupling between the pseudogap and superconductivity, imposing stringent constraints on microscopic theories of high-temperature superconductivity.

## Introduction

A continuing conundrum in cuprate superconductors concerns the nature of the anomalous normal state from which superconductivity develops[1–3]. The pseudogap manifests as a partial depletion of low-energy spectral weight well above the superconducting transition temperature[4,5]. After decades of investigation, the physical origin of the pseudogap remains intensely debated: it is predominantly interpreted either as a precursor state of preformed Cooper pairs lacking global phase coherence[6–8], or as a distinct competing order[9], such as intertwined density waves[10,11]. Delineating the precise thermodynamic crossovers and the phase diagram of these states is essential for elucidating the microscopic pairing mechanism. To date, experimental efforts to map their interplay have predominantly relied on chemical doping[1,2]. However, compositional tuning inevitably introduces spatial disorder and impurity scattering, smearing the intrinsic phase boundaries and obscuring the true relationship between the pseudogap and superconductivity[12,13].

Hydrostatic pressure serves as a pristine tuning parameter that avoids the lattice disorder introduced by chemical substitution. Continuous volumetric compression at a fixed stoichiometry renormalizes the inter-site hopping integrals and on-site Coulomb interactions, thereby dictating the balance between strong Mott correlations and charge itinerancy in correlated systems[14]. This approach has reshaped the phase diagrams of diverse quantum materials, ranging from stabilizing superconductivity in hydrides[15,16] and nickelates[17,18] to modulating the superconducting domes of cuprates[19–21] and iron pnictides[22,23]. While high-pressure electrical transport measurements successfully trace the phase transitions, they lack the spectroscopic sensitivity required to quantify the underlying energy gaps associated with these emergent states.

Directly probing the gaps under compression faces experimental challenges: the restricted optical access and strong multiphonon absorption of the diamond anvils obscure the subtle low-energy excitations in conventional spectroscopy[24]. Consequently, the investigations of cuprate energy gaps have been confined to modest pressures[20,25,26], leaving their evolution across the extended phase diagram unexplored. Ultrafast optical spectroscopy circumvents these limitations by coupling femtosecond pulses into the high-pressure cavity[27]. By measuring the transient reflectivity, the quasiparticle relaxation channels can be disentangled in time domain, isolating the dynamics of distinct electronic phases[28,29]. Recently, the high-pressure ultrafast platform unveiled the temperature-dependent gap dynamics of nickelate superconductors[30,31].

Here, we employ ultrafast optical spectroscopy to investigate the evolution of the characteristic energy scales in slightly underdoped $Bi_2Sr_2CaCu_2O_{8+\delta}$ (Bi2212) up to 37 GPa. Through decomposition and fitting the multi-component transient reflectivity, we simultaneously extract the transition temperatures ($T^*$, $T_c$) and their corresponding energy gaps ($\Delta_{PG}$, $\Delta_{SC}$). Strikingly, the pseudogap behavior breaks the proportional scaling established by chemical doping: $T^*$ rises monotonically under compression, whereas $\Delta_{PG}$ is initially attenuated before plateauing at a finite value above 23 GPa. In contrast, $T_c$ and $\Delta_{SC}$ exhibit a synchronized dome-like evolution before the superconducting state is extinguished at 37 GPa. Within the superconducting dome, a sharp collapse of the coupling ratio $2\Delta_{\mathrm{SC}}/k_B T_c$ signifies a pressure-driven dimensional crossover, quenching two-dimensional phase fluctuations to stabilize global three-dimensional coherence. The decoupled pressure evolution provides a solid experimental foundation favoring the two-gap scenario and demonstrates the distinct microscopic origins for the two phases.

**Results**

Measurements of AC magnetic susceptibility on the Bi2212 single crystal exhibit a sharp superconducting transition at an onset $T_C \approx 91.5$ K with a narrow transition width (Extended Data Fig. 1), indicating a slightly underdoped level. To establish an ambient-pressure baseline for subsequent compression experiments, the crystal was loaded into a diamond anvil cell (DAC, Fig. **1a**), and temperature-dependent differential reflectivity ( $\Delta R/R$ ) was recorded using a two-color pump-probe configuration (Extended Data Fig. 2). The measured transients display multi-component relaxation dynamics (Fig. **1b**), which we model using a standard bi-exponential decay function[28]: $\Delta R/\mathrm{R}(t) = A_1 e^{-t/\tau_1} + A_2 e^{-t/\tau_2} + C$. Here, $A_i$ and $\tau_i$ denote the amplitude and relaxation time of the respective channels, and $C$ represents a long-lived offset. This decomposition isolates two relaxation pathways. A fast, positive component ($A_1$) emerging below $T^*$ is assigned to the pseudogap[32,33]. While a slower, negative decay ($A_2$) persists at all temperatures, its magnitude undergoes a sharp enhancement below $T_c$ due to quasiparticle recombination across the superconducting gap[34,35], reducing to a weak normal-state metallic response at higher temperatures. The observed ambient-pressure dynamics agree with reported studies on Bi2212[32,33,36,37], confirming that the DAC assembly preserves the intrinsic transient response.

The temperature-dependent $\Delta R/R$ traces measured at selected pressures are plotted as false-color maps (Fig. **2**). At near-ambient pressure, the positive pseudogap domain (red) visually separates a weak negative normal-state response at high temperatures from the pronounced negative superconducting signal at low temperatures (blue). Under compression, the positive pseudogap region expands progressively, replacing the high-temperature negative metallic response. By 14 GPa, the positive domain extends beyond our experimental limit of 300 K. Meanwhile, the low-temperature superconducting signal exhibits a non-monotonic pressure dependence, initially shifting to higher temperatures before being severely suppressed. At 37 GPa, the negative superconducting signature vanishes, yielding a positive transient profile across all measured temperatures (Figs. **1c** and **2f**).

Applying the temporal decomposition across the entire pressure range, we extract the temperature-dependent amplitudes (Fig. 3 and Extended Data Fig. 4) and relaxation lifetimes (Extended Data Fig. 5) of the two relaxation channels. As temperature increases, the positive amplitude $A_1$ diminishes, and the temperature at which this channel vanishes defines the pseudogap thermal onset $T^*$[35]. Concurrently, the negative amplitude $A_2$ collapses near $T_c$. At lower pressures, the rapid growth of $A_2$ below $T_c$ depletes $A_1$, indicating a competitive spectral weight transfer[32,37,38]. The spectroscopic determination of $T_c$ is further validated by the divergence of the relaxation lifetime $\tau_2$ near the transition due to the phonon-bottleneck effect[28,33].

Tracking these characteristic temperatures constructs the pressure-temperature phase diagram (Fig. **4a**), revealing a separation of the two electronic orders: while $T^*$ increases quasi-linearly from 170 K near ambient pressure to exceed 300 K beyond 14 GPa, $T_c$ exhibits a non-monotonic dome-like dependence that peaks near 6 GPa before fully suppressed between 29 and 37 GPa.

To extract the energy scales of the pseudogap and superconducting phases, we employ the phenomenological Rothwarf-Taylor (RT) model to analyze the temperature-dependent transient amplitudes (Fig. **3**)[34,35]. In gapped systems, quasiparticle recombination is impeded by the continuous reabsorption of the emitted high-frequency phonons ($\hbar\omega \geq 2\Delta$). The cyclic energy exchange inherently prolongs the effective quasiparticle lifetime, manifesting as phonon bottleneck effect[28]. In the weak-excitation limit, the transient reflectivity amplitudes scale linearly with the photoinduced quasiparticle density, which is in turn governed by the thermally excited quasiparticle population. Consequently, the non-equilibrium amplitudes for a temperature-independent pseudogap and a temperature-dependent superconducting gap are expressed as[35]:

$$A_{PG}\,(T) \propto \frac{\Phi\,/\,\Delta}{1+\gamma\,\exp[-\Delta/k_B T]} \tag{1}$$

$$A_{SC}(T) \propto \frac{\Phi/(\Delta(T)+k_B T/2)}{1+\gamma\sqrt{2k_B T/\Delta(T)}\,\exp[-\Delta(T)/k_B T]} \tag{2}$$

Here, $\Phi$ is the pump fluence, $k_B$ is the Boltzmann constant, and $\gamma$ represents the ratio of phonon and quasiparticle densities of states. To capture the thermal evolution of the superconducting condensate, we apply the standard BCS interpolation[39] $\Delta_{SC}(T) \approx \Delta_{SC}(0)\tanh\left(1.74\sqrt{T_c/\mathrm{T}-1}\right)$. Equations **(1)** and **(2)** accurately reproduce the amplitude evolution up to 23 GPa. At 29 GPa, the superconducting recombination time diverges due to the phonon bottleneck effect. Consequently, the residual negative decay evolves into a constant offset, which serves as the effective amplitude to evaluate the superconducting gap. Approaching the high-pressure boundary of superconductivity, a faster positive decay component emerges, potentially driven by the electronic reconstruction of the ground state[21,40]. The slower positive transient retains a relaxation lifetime consistent with the low-pressure pseudogap dynamics, thereby enabling the continuous extraction of the pseudogap. The overall pressure evolution of these gapped states is summarized in Fig. **4b**. At ambient pressure, the extracted pseudogap energy ($\Delta_{\mathrm{PG}} \approx$ 72 meV) and zero-temperature superconducting gap ($\Delta_{\mathrm{SC}} \approx$ 38 meV) quantitatively match established literature values for near-optimally doped Bi2212[41]. Under compression, $\Delta_{\mathrm{PG}}$ undergoes a steep initial reduction before increasing slightly above 23 GPa, persisting up to the highest investigated pressure of 37 GPa. In contrast, $\Delta_{\mathrm{SC}}$ remains nearly constant up to 6 GPa, beyond which it rapidly collapses and vanishes entirely above 29 GPa.

## Discussion

Our ultrafast spectroscopic measurements establish a comprehensive high-pressure phase diagram for Bi2212 up to 37 GPa, systematically extracting the thermal onset temperatures and their corresponding energy scales of both the pseudogap and superconducting states. Phenomenologically, the concurrent rise of $T^*$ and suppression of $T_c$ resembles the conventional phase trajectory of the deeply underdoped regime[1,2]. However, our high-pressure thermoelectric measurements rule out this scenario. The pressure-driven evolution of the Seebeck coefficient, measured up to 9 GPa (Extended Data Fig. 6), confirms a monotonic increase in the in-plane hole concentration ($p$), suggesting a progressive pressure-induced charge transfer from the reservoir spacer layers into the $CuO_2$ planes[21,42–44]. Therefore, the simultaneous evolution of $T^*$ and $T_c$ alongside increasing $p$ explicitly violates the empirical phase diagram established via chemical doping. Furthermore, while $T^*$ rises steadily beyond 14 GPa, the pseudogap energy $\Delta_{\mathrm{PG}}$ is continuously suppressed. This inverse scaling contrasts with the superconducting state, where $T_c$ and $\Delta_{SC}$ exhibit synchronized evolution. The decoupling of the energy gap from its onset temperature demonstrates that these two characteristic scales of the pseudogap emerge from distinct microscopic origins.

The monotonic elevation of $T^*$ manifests the pressure-driven reinforcement of short-range antiferromagnetic correlations. Hydrostatic compression contracts the in-plane Cu-O bond lengths, amplifying the nearest-neighbor electronic hopping integral $t$[45,40,46]. Because the on-site Coulomb repulsion $U$ exhibits a much weaker pressure dependence, the kinetic energy gain drives an enhancement of the planar superexchange interaction, $J \approx 4t^2/\mathrm{U}$. This magnetic exchange hardening is corroborated by the high-energy blueshift of two-magnon excitations resolved in high-pressure Raman spectroscopy[47,48]. Concurrently, the broadened electronic bandwidth $W$ attenuates the Mott correlation strength $U/W$ [1,49]. The ensuing delocalization promotes charge itinerancy, progressively suppressing the single-particle excitation gap $\Delta_{\mathrm{PG}}$. Above 23 GPa, $\Delta_{\mathrm{PG}}$ increases slightly, to approximately 34 meV at 37 GPa. This anomaly coincides with an isostructural lattice stiffening[50], wherein charge redistribution between the BiO and $CuO_2$ layers revives the local Coulomb repulsion $U$ to stabilize the effective $U/W$ ratio[40]. These disparate pressure dependencies unveil an underlying spin-charge decoupling within the pseudogap regime: $T^*$ is dictated by the magnetic exchange scale $J$, whereas $\Delta_{\mathrm{PG}}$ is governed by the charge localization scale $U/W$.

By simultaneously tracking $T_c$ and $\Delta_{\mathrm{SC}}$, we evaluate the effective coupling ratio $2\Delta_{\mathrm{SC}}/k_B T_c$ to quantify the pressure-driven evolution of the superconducting condensate (Fig. **4c**). Below 6 GPa, the ratio plateaus near $\sim$10, a hallmark of phase-fluctuating superconductivity where $T_c$ is constrained well below the mean-field pairing value[6]. Between 6 and 11 GPa, however, the ratio sharply declines, approaching the weak-coupling $d$-wave limit ($\approx$4.3) under further compression[51,52]. The steep reduction indicates a two-to-three-dimensional superconducting crossover, triggered by the suppression of $\Delta_{\mathrm{PG}}$. In momentum space, the interlayer hopping matrix element $t_{\perp}(\mathbf{k})$ is maximal at the antinodes. Because the pseudogap predominantly depletes the density of states in these specific regions[53], its suppression restores the low-energy spectral weight required for interlayer tunneling. Coupled with the reduced interplane spacing, the restored antinodal coherence enhances the out-of-plane superfluid phase stiffness, quenching two-dimensional phase fluctuations and driving the system into a three-dimensional superconducting state. Crucially, the critical pressure aligns with both the onset of global $c$-axis transport[54] near 8 GPa in hole-doped Bi2212 and the Fermi surface reconstruction[55] in electron-doped cuprates, revealing a generic boundary for electronic instabilities spanning the cuprate phase diagram.

A central implication of our extended phase diagram is the decoupling of the pseudogap from the superconducting state. Unlike chemical underdoping, where phase fluctuations suppress $T_c$, or overdoping, where carrier delocalization destroys both the pairing amplitude and the pseudogap, high pressure quenches superconductivity via an amplitude collapse of $\Delta_{SC}$ within a three-dimensional coherent state. If the pseudogap were a precursor state of incoherent Cooper pairs[6,7], it would inevitably vanish alongside the amplitude collapse[6,8]. Instead, the pseudogap persists beyond the disappearance of superconductivity at 37 GPa, providing direct evidence for the two-gap scenario. By establishing the pseudogap as a distinct electronic order rather than a fluctuating pairing state, our findings place stringent constraints on microscopic theories of high-temperature superconductivity.

## Method

**Sample fabrication and characterization:** High-quality $Bi_2Sr_2CaCu_2O_{8+\delta}$ single crystals were grown using the traveling-solvent floating-zone technique[56]. Bulk superconductivity was confirmed using AC magnetic susceptibility, revealing a sharp superconducting transition with an onset temperature of 91.5 K and a midpoint of 89.5 K (Extended Data Fig. 1). Freshly cleaved samples from this characterized batch were utilized for all subsequent high-pressure ultrafast optical measurements.

**High-pressure ultrafast optical spectroscopy:** High-pressure environments were generated using a non-magnetic beryllium-copper diamond anvil cell (DAC) equipped with 500-$\mu$m culet anvils. The Bi2212 crystal was mounted within a pre-indented rhenium gasket, with KBr powder acting as a quasi-hydrostatic pressure-transmitting medium. *In situ* pressure calibration was performed by monitoring the ruby $R_1$ fluorescence line shift. The DAC was integrated into a continuous-flow liquid-helium cryostat operating between 20 and 300 K.

Two-color pump-probe spectroscopy was conducted using a mode-locked Yb:KGW amplifier coupled with an optical parametric amplifier, generating 50-fs pulses at a 50 kHz repetition rate. Non-equilibrium quasiparticles were photoexcited by 400 nm (3.10 eV) pump pulses and their subsequent relaxation dynamics were probed by the 800 nm (1.55 eV) fundamental beam. The pump and probe beams were collinearly combined via a dichroic mirror and focused onto the sample through a 5× objective, yielding focal spot diameters of 34 $\mu$m and 14 $\mu$m, respectively (Extended Data Fig. 2). To ensure operation within the weak-perturbation limit, the incident pump and probe fluences were maintained at 60 $\mu J/cm^2$ and 11 $\mu J/cm^2$, respectively. The transient reflectivity ($\Delta R/R$) was measured in a reflection geometry via balanced photodetection synchronized with a mechanical chopper and lock-in amplification.

## Acknowledgements

This work was supported by the National Key Research and Development Program of China (Grant Nos. 2024YFA1611300, 2023YFA1608900, 2025YFA1411500, 2025YFE0202100, and 2021YFA1401800), the National Natural Science Foundation of China (Grant Nos. 12375304, 12404163, 12574349, and 12488201). E.E.M.C. acknowledges support from the Singapore Ministry of Education (MOE) Academic Research Fund Tier 3 (MOE-MOET32023-0003) grant. This work was carried out at the Synergetic Extreme Condition User Facility (SECUF, https://cstr.cn/31123.02.SECUF).

## Author contributions

L.Z., X.Y., and X.W. conceived the project. Y.M. performed the ultrafast optical experiments. W.M. grew and characterized the single crystals. L.C. performed the Seebeck coefficient measurements. Y.M., X.Y. and X.W. analyzed the data and wrote the manuscript. All authors discussed the results and commented on the manuscript.

## Competing interests

The authors declare no competing interests.

## Data availability

Source data are provided with this paper. All other data that support the findings of this study are available from the corresponding author upon reasonable request.

## Reference

1. Lee, P. A., Nagaosa, N. & Wen, X.-G. Doping a Mott insulator: Physics of high-temperature superconductivity. *Rev. Mod. Phys.* **78**, 17–85 (2006).

2. Keimer, B., Kivelson, S. A., Norman, M. R., Uchida, S. & Zaanen, J. From quantum matter to high-temperature superconductivity in copper oxides. *Nature* **518**, 179–186 (2015).

3. Proust, C. & Taillefer, L. The Remarkable Underlying Ground States of Cuprate Superconductors. *Annu. Rev. Condens. Matter Phys.* **10**, 409–429 (2019).

4. Timusk, T. & Statt, B. The pseudogap in high-temperature superconductors: an experimental survey. *Rep. Prog. Phys.* **62**, 61–122 (1999).

5. Norman, M. R., Pines, D. & Kallin, C. The pseudogap: friend or foe of high $T_c$? *Adv. Phys.* **54**, 715–733 (2005).

6. Emery, V. J. & Kivelson, S. A. Importance of phase fluctuations in superconductors with small superfluid density. *Nature* **374**, 434–437 (1995).

7. Corson, J., Mallozzi, R., Orenstein, J., Eckstein, J. N. & Bozovic, I. Vanishing of phase coherence in underdoped $Bi_2Sr_2CaCu_2O_{8+\delta}$. *Nature* **398**, 221–223 (1999).

8. Chen, Q., Wang, Z., Boyack, R., Yang, S. & Levin, K. When superconductivity crosses over: From BCS to BEC. *Rev. Mod. Phys.* **96**, 025002 (2024).

9. Fradkin, E., Kivelson, S. A. & Tranquada, J. M. *Colloquium*: Theory of intertwined orders in high temperature superconductors. *Rev. Mod. Phys.* **87**, 457–482 (2015).

10. Ghiringhelli, G. *et al.* Long-Range Incommensurate Charge Fluctuations in $(Y,Nd)Ba_2Cu_3O_{6+x}$. *Science* **337**, 821–825 (2012).

11. Chang, J. *et al.* Direct observation of competition between superconductivity and charge density wave order in $YBa_2Cu_3O_{6.67}$. *Nat. Phys.* **8**, 871–876 (2012).

12. Eisaki, H. *et al.* Effect of chemical inhomogeneity in bismuth-based copper oxide superconductors. *Phys. Rev. B* **69**, 064512 (2004).

13. Nie, L., Tarjus, G. & Kivelson, S. A. Quenched disorder and vestigial nematicity in the pseudogap regime of the cuprates. *Proc. Natl. Acad. Sci. U.S.A.* **111**, 7980–7985 (2014).

14. Imada, M., Fujimori, A. & Tokura, Y. Metal-insulator transitions. *Rev. Mod. Phys.* **70**, 1039–1263 (1998).

15. Drozdov, A. P., Eremets, M. I., Troyan, I. A., Ksenofontov, V. & Shylin, S. I. Conventional superconductivity at 203 kelvin at high pressures in the sulfur hydride system. *Nature* **525**, 73–76 (2015).

16. Drozdov, A. P. *et al.* Superconductivity at 250 K in lanthanum hydride under high pressures. *Nature* **569**, 528–531 (2019).

17. Sun, H. *et al.* Signatures of superconductivity near 80 K in a nickelate under high pressure. *Nature* **621**, 493–498 (2023).

18. Zhang, Y. *et al.* High-temperature superconductivity with zero resistance and strange-metal behaviour in $La_3Ni_2O_{7-\delta}$. *Nat. Phys.* **20**, 1269–1273 (2024).

19. Gao, L. *et al.* Superconductivity up to 164 K in $HgBa_2Ca_{m-1}Cu_mO_{2m+2+\delta}$ (m = 1, 2, and 3) under quasihydrostatic pressures. *Phys. Rev. B* **50**, 4260–4263 (1994).

20. Chen, X.-J. *et al.* Enhancement of superconductivity by pressure-driven competition in electronic order. *Nature* **466**, 950–953 (2010).

21. Zhou, Y. *et al.* Quantum phase transition from superconducting to insulating-like state in a pressurized cuprate superconductor. *Nat. Phys.* **18**, 406–410 (2022).

22. Takahashi, H. *et al.* Superconductivity at 43 K in an iron-based layered compound $LaO_{1-x}F_xFeAs$. *Nature* **453**, 376–378 (2008).

23. Torikachvili, M. S., Bud'ko, S. L., Ni, N. & Canfield, P. C. Pressure Induced Superconductivity in $CaFe_2As_2$. *Phys. Rev. Lett.* **101**, 057006 (2008).

24. Mao, H.-K., Chen, X.-J., Ding, Y., Li, B. & Wang, L. Solids, liquids, and gases under high pressure. *Rev. Mod. Phys.* **90**, 015007 (2018).

25. Souliou, S. M. *et al.* Rapid suppression of the charge density wave in $YBa_2Cu_3O_{6.6}$

under hydrostatic pressure. *Phys. Rev. B* **97**, 020503 (2018).

26. Vinograd, I. *et al.* Nuclear magnetic resonance study of charge density waves under hydrostatic pressure in $YBa_2Cu_3O_y$. *Phys. Rev. B* **100**, 094502 (2019).

27. Wu, Y. L. *et al.* *On-site in situ* high-pressure ultrafast pump–probe spectroscopy instrument. *Rev. Sci. Instrum.* **92**, 113002 (2021).

28. Giannetti, C. *et al.* Ultrafast optical spectroscopy of strongly correlated materials and high-temperature superconductors: a non-equilibrium approach. *Adv. Phys.* **65**, 58–238 (2016).

29. de la Torre, A. *et al.* Colloquium: Nonthermal pathways to ultrafast control in quantum materials. *Rev. Mod. Phys.* **93**, 041002 (2021).

30. Meng, Y. *et al.* Density-wave-like gap evolution in $La_3Ni_2O_7$ under high pressure revealed by ultrafast optical spectroscopy. *Nat. Commun.* **15**, 10408 (2024).

31. Xu, S. *et al.* Collapse of density wave and emergence of superconductivity in pressurized-$La_4Ni_3O_{10}$ evidenced by ultrafast spectroscopy. *Nat. Commun.* **16**, 7039 (2025).

32. Liu, Y. H. *et al.* Direct Observation of the Coexistence of the Pseudogap and Superconducting Quasiparticles in $Bi_2Sr_2CaCu_2O_{8+y}$ by Time-Resolved Optical Spectroscopy. *Phys. Rev. Lett.* **101**, 137003 (2008).

33. Toda, Y. *et al.* Quasiparticle relaxation dynamics in underdoped $Bi_2Sr_2CaCu_2O_{8+\delta}$ by two-color pump-probe spectroscopy. *Phys. Rev. B* **84**, 174516 (2011).

34. Rothwarf, A. & Taylor, B. N. Measurement of Recombination Lifetimes in Superconductors. *Phys. Rev. Lett.* **19**, 27–30 (1967).

35. Kabanov, V. V., Demsar, J., Podobnik, B. & Mihailovic, D. Quasiparticle relaxation dynamics in superconductors with different gap structures: Theory and experiments on $YBa_2Cu_3O_{7-\delta}$. *Phys. Rev. B* **59**, 1497–1506 (1999).

36. Nair, S. K. *et al.* Quasiparticle dynamics in overdoped $Bi_{1.4}Pb_{0.7}Sr_{1.9}CaCu_2O_{8+\delta}$: Coexistence of superconducting gap and pseudogap below $T_c$. *Phys. Rev. B* **82**, 212503

(2010).

37. Coslovich, G. *et al.* Competition Between the Pseudogap and Superconducting States of $Bi_2Sr_2Ca_{0.92}Y_{0.08}Cu_2O_{8+\delta}$ Single Crystals Revealed by Ultrafast Broadband Optical Reflectivity. *Phys. Rev. Lett.* **110**, 107003 (2013).

38. Chia, E. E. M. *et al.* Observation of Competing Order in a High-$T_c$ Superconductor Using Femtosecond Optical Pulses. *Phys. Rev. Lett.* **99**, 147008 (2007).

39. Demsar, J., Podobnik, B., Kabanov, V. V., Wolf, Th. & Mihailovic, D. Superconducting Gap $\Delta_c$, the Pseudogap $\Delta_p$, and Pair Fluctuations above $T_c$ in Overdoped $Y_{1-x}Ca_xBa_2Cu_3O_{7-\delta}$ from Femtosecond Time-Domain Spectroscopy. *Phys. Rev. Lett.* **82**, 4918–4921 (1999).

40. Du, X., Zhang, J.-F., Lu, Z.-Y. & Liu, K. Origin of insulating-like behavior of $Bi_2Sr_2CaCu_2O_{8+x}$ under pressure: A first-principles study. *Phys. Rev. B* **112**, 045113 (2025).

41. Hashimoto, M., Vishik, I. M., He, R.-H., Devereaux, T. P. & Shen, Z.-X. Energy gaps in high-transition-temperature cuprate superconductors. *Nat. Phys.* **10**, 483–495 (2014).

42. Jorgensen, J. D. *et al.* Pressure-induced charge transfer and $dT_c/dP$ in $YBa_2Cu_3O_{7-x}$. *Physica C* **171**, 93–102 (1990).

43. Ambrosch-Draxl, C., Sherman, E. Ya., Auer, H. & Thonhauser, T. Pressure-Induced Hole Doping of the Hg-Based Cuprate Superconductors. *Phys. Rev. Lett.* **92**, 187004 (2004).

44. Gourgout, A. *et al.* Effect of pressure on the pseudogap and charge density wave phases of the cuprate Nd-LSCO probed by thermopower measurements. *Phys. Rev. Res.* **3**, 023066 (2021).

45. Pavarini, E., Dasgupta, I., Saha-Dasgupta, T., Jepsen, O. & Andersen, O. K. Band-Structure Trend in Hole-Doped Cuprates and Correlation with $T_{c\,max}$. *Phys. Rev. Lett.* **87**, 047003 (2001).

46. Markiewicz, R. S., Sahrakorpi, S., Lindroos, M., Lin, H. & Bansil, A. One-band tight-binding model parametrization of the high-$T_c$ cuprates including the effect of $k_z$ dispersion. *Phys. Rev. B* **72**, 054519 (2005).

47. Xin, J. *et al.* Lattice Effect on the Superexchange Interaction in Antiferromagnetic $Bi_{2.1}Sr_{1.9}CaCu_2O_{8+\delta}$. *J. Phys. Chem. C* **128**, 7223–7234 (2024).

48. Plakida, N. *High-Temperature Cuprate Superconductors: Experiment, Theory, and Applications*. (Springer, Berlin, Heidelberg, 2010). doi:10.1007/978-3-642-12633-8.

49. Phillips, P. Mottness. *Ann. Phys.* **321**, 1634–1650 (2006).

50. Mark, A. C. *et al.* Structure and equation of state of $Bi_2Sr_2Ca_{n-1}Cu_nO_{2n+4+\delta}$ from x-ray diffraction to megabar pressures. *Phys. Rev. Mater.* **7**, 064803 (2023).

51. Won, H. & Maki, K. *d*-wave superconductor as a model of high-$T_c$ superconductors. *Phys. Rev. B* **49**, 1397–1402 (1994).

52. Scalapino, D. J. The case for $d_{x^2-y^2}$ pairing in the cuprate superconductors. *Phys. Rep.* **250**, 329 (1995).

53. Ioffe, L. B. & Millis, A. J. Zone-diagonal-dominated transport in high-$T_c$ cuprates. *Phys. Rev. B* **58**, 11631–11637 (1998).

54. Guo, J. *et al.* Crossover from two-dimensional to three-dimensional superconducting states in bismuth-based cuprate superconductor. *Nat. Phys.* **16**, 295–300 (2020).

55. Zhao, J. *et al.* Pressure-induced Lifshitz and quantum phase transitions in electron-doped cuprate superconductors. *Proc. Natl. Acad. Sci.* **123**, e2622416123 (2026).

56. Gu, G. D., Takamuku, K., Koshizuka, N. & Tanaka, S. Large single crystal Bi-2212 along the c-axis prepared by floating zone method. *J. Cryst. Growth* **130**, 325–329 (1993).

57. Cusack, N. & Kendall, P. The absolute scale of thermoelectric power at high temperature. *Proc. Phys. Soc.* **72**, 898–901 (1958).

58. Obertelli, S. D., Cooper, J. R. & Tallon, J. L. Systematics in the thermoelectric power of high-$T_c$ oxides. *Phys. Rev. B* **46**, 14928–14931 (1992).

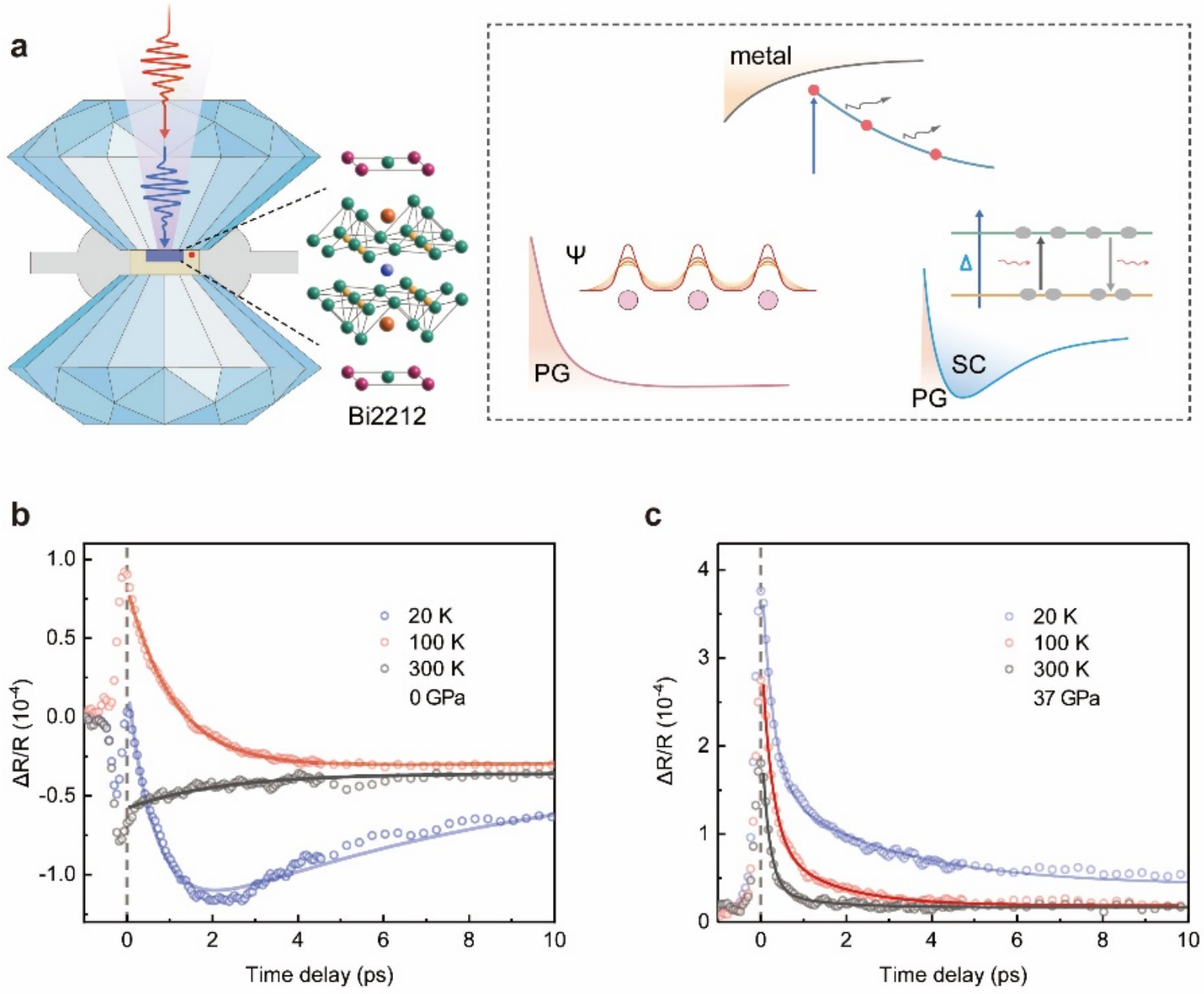

**Fig. 1 | High-pressure ultrafast optical spectroscopy and non-equilibrium quasiparticle dynamics. a**, Schematic of the diamond anvil cell setup for two-color pump-probe measurements (left), alongside the phenomenological models of the transient optical signatures for the normal, pseudogap, and superconducting phases (right). **b,** Temperature-dependent transient reflectivity ($\Delta R/R$) traces near ambient pressure. The profiles at 20, 100 and 300 K resolve the corresponding relaxation dynamics of the superconducting, pseudogap and normal metallic states, respectively. **c,** Transient reflectivity at 37 GPa. The negative superconducting response vanishes at 20 K, replaced by a positive signal. Solid lines represent bi-exponential fits, which is restricted to delay times $t > 0$ ps (vertical dashed line) to exclude coherent artifacts.

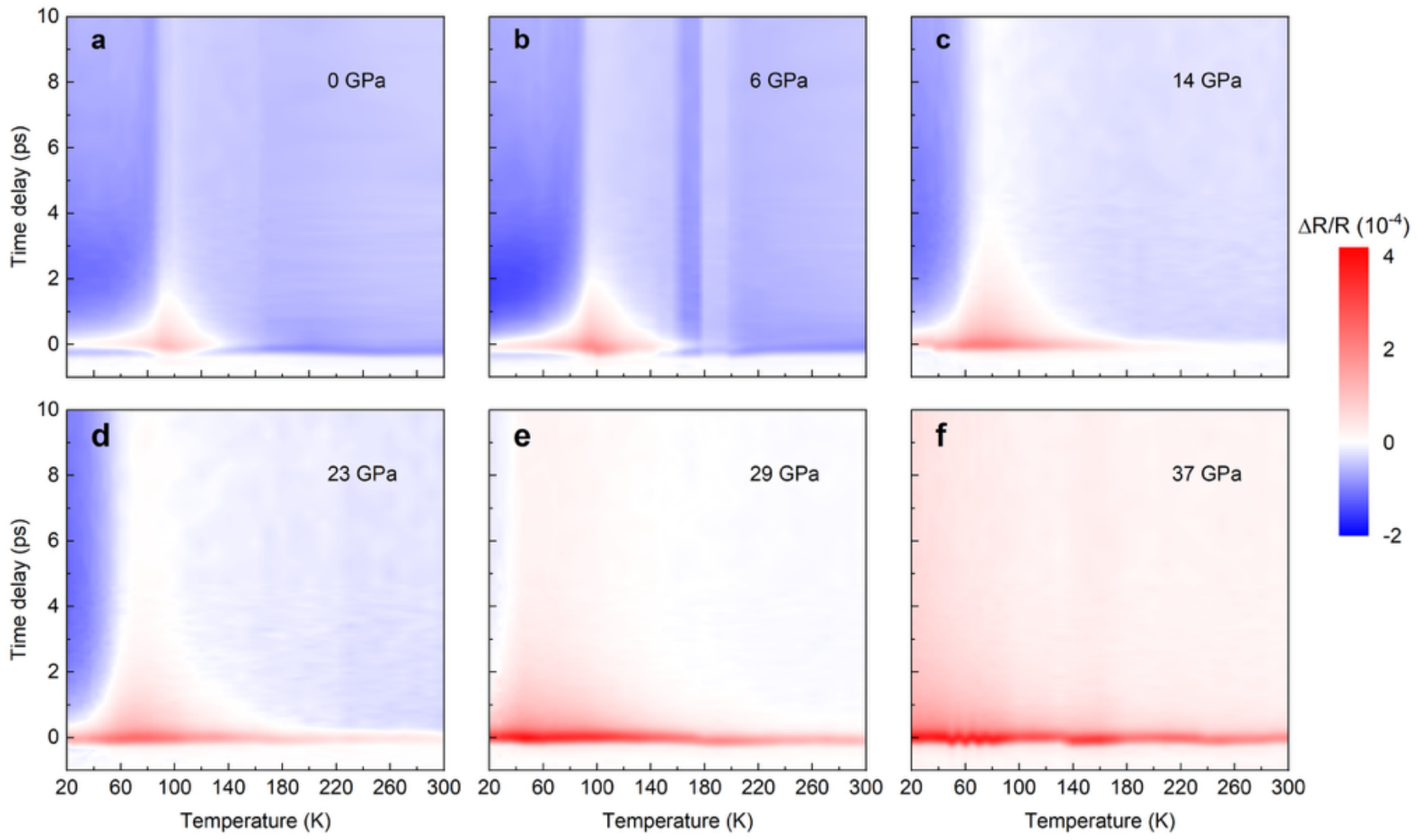

**Fig. 2 | Pressure evolution of the non-equilibrium quasiparticle dynamics. a-f,** False-color maps of the transient reflectivity ($\Delta R/R$) versus pump-probe delay and temperature at selected pressures. The superconducting condensate manifests as a negative (blue) response, whereas the pseudogap dominates the initial positive (red) transient.

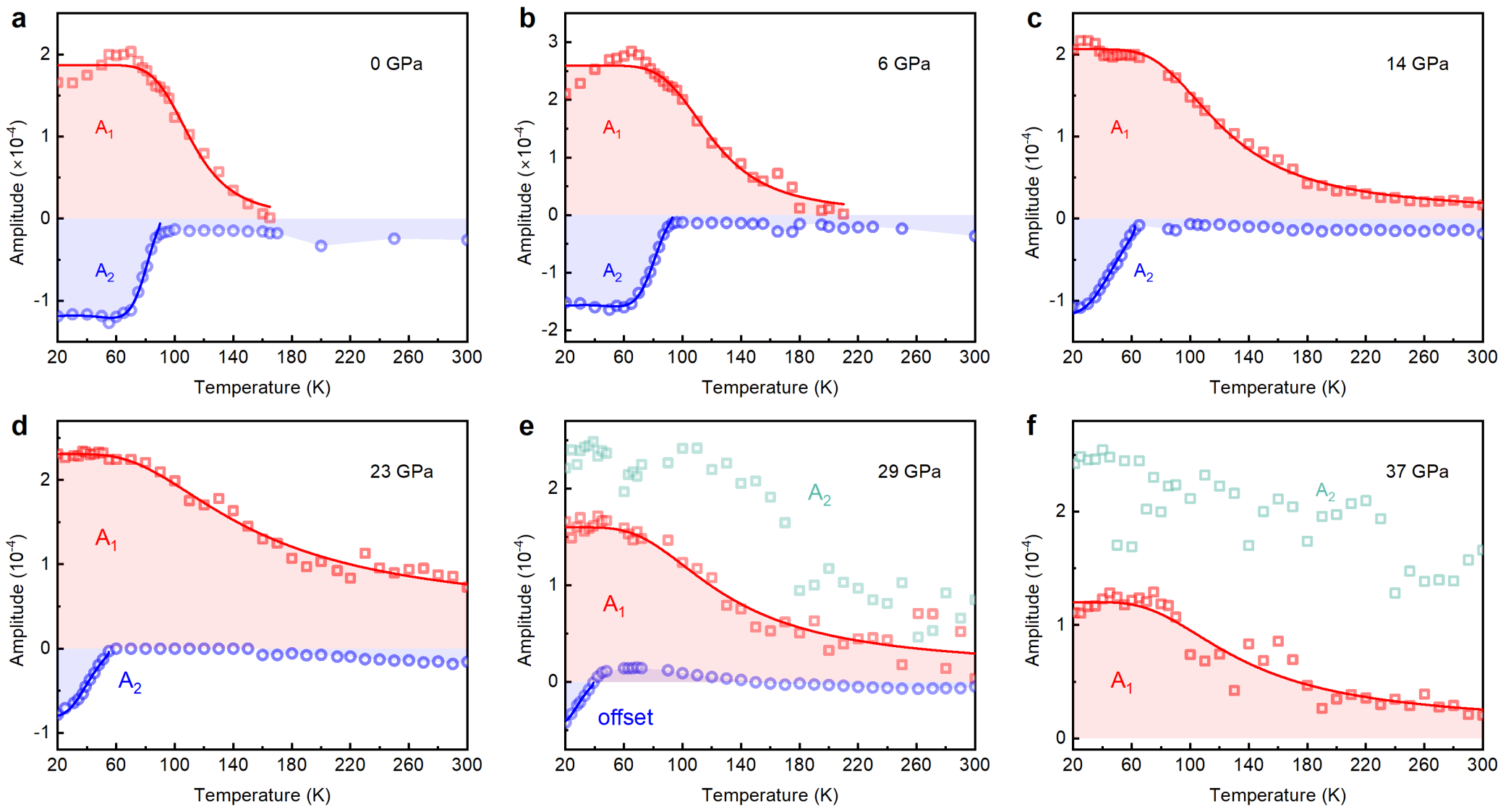

**Fig. 3 | Temperature-dependent excitation amplitudes and Rothwarf-Taylor analysis. a-d**, Temperature evolution of the amplitudes extracted using a unified bi-exponential decay model at representative pressures below 23 GPa. **e**, Amplitude profile at 29 GPa. The divergence of the superconducting recombination time transforms the negative superconducting response into a long-lived negative offset. **f**, Amplitudes at 37 GPa. The complete suppression of the superconducting response indicates the transition to an insulating-like phase, leaving only the positive decay channels. Solid lines represent the fits to the phenomenological Rothwarf-Taylor (RT) model.

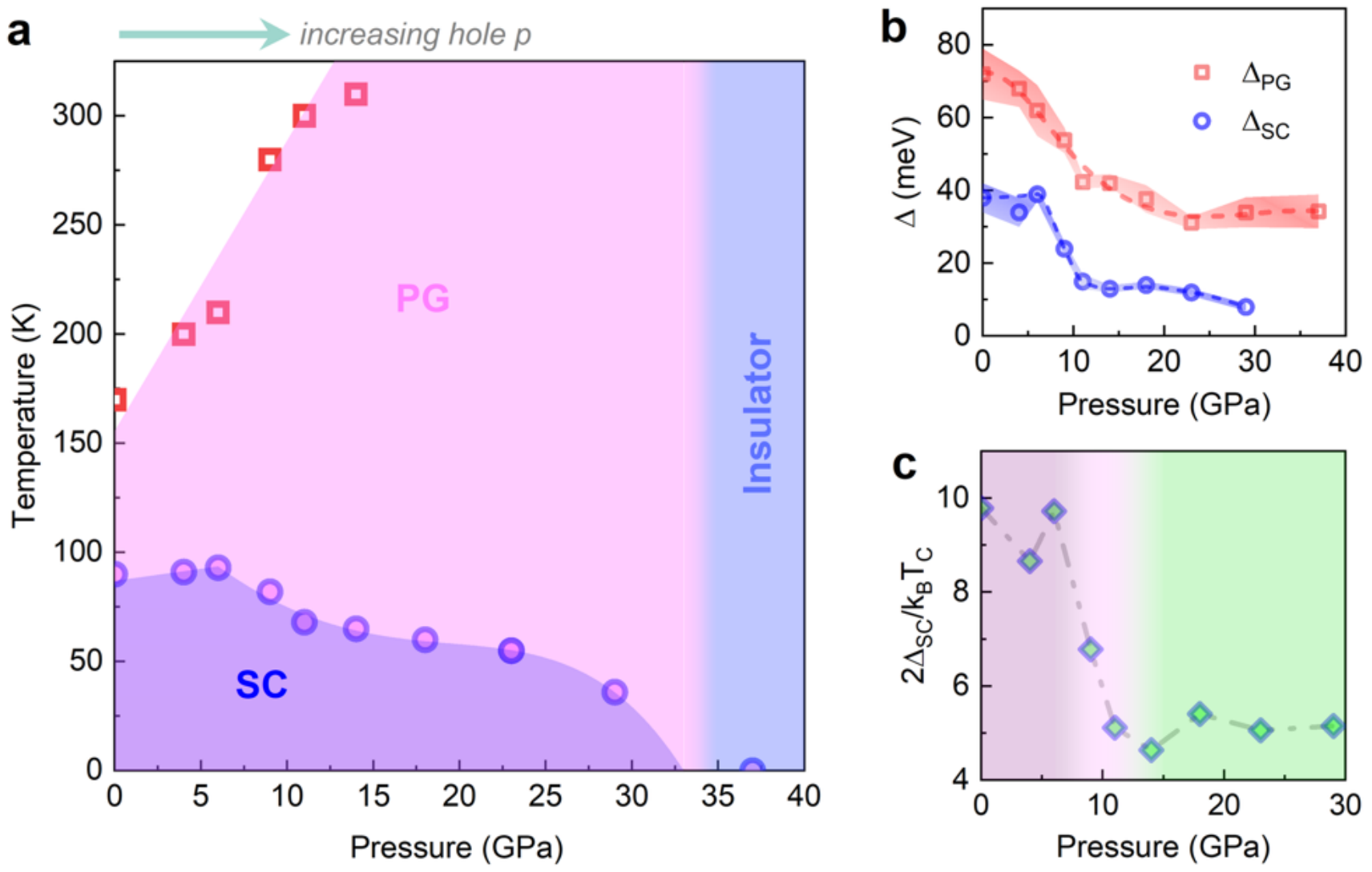

**Fig. 4 | High-pressure electronic phase diagram and energy-scale decoupling. a**, Pressure-temperature phase diagram. The superconducting transition $T_c$ exhibits a dome-like trajectory, peaking at 98 K prior to its complete suppression between 29 and 37 GPa. Conversely, the pseudogap thermal onset $T^*$ increases monotonically, exceeding room temperature beyond 14 GPa. The top arrow denotes the pressure-induced enhancement of the effective hole concentration ($p$). **b**, Pressure evolution of the extracted energy scales. The superconducting gap $\Delta_{\mathrm{SC}}$ tracks the $T_c$ dome. In contrast, the pseudogap energy $\Delta_{\mathrm{PG}}$ decreases continuously up to 23 GPa before increasing slightly to ~34 meV at 37 GPa, revealing an anomalous decoupling between $T^*$ and $\Delta_{\mathrm{PG}}$. Shaded bands represent the standard error of the fits. **c**, Effective coupling strength and dimensional crossover. The ratio $2\Delta_{SC}/k_BT_c$ plateaus near ~10 below 6 GPa before declining towards the weak-coupling BCS limit (~4.3) at higher pressures. This step-like reduction represents a crossover from a fluctuation-dominated two-dimensional regime to a globally coherent three-dimensional superconducting state.

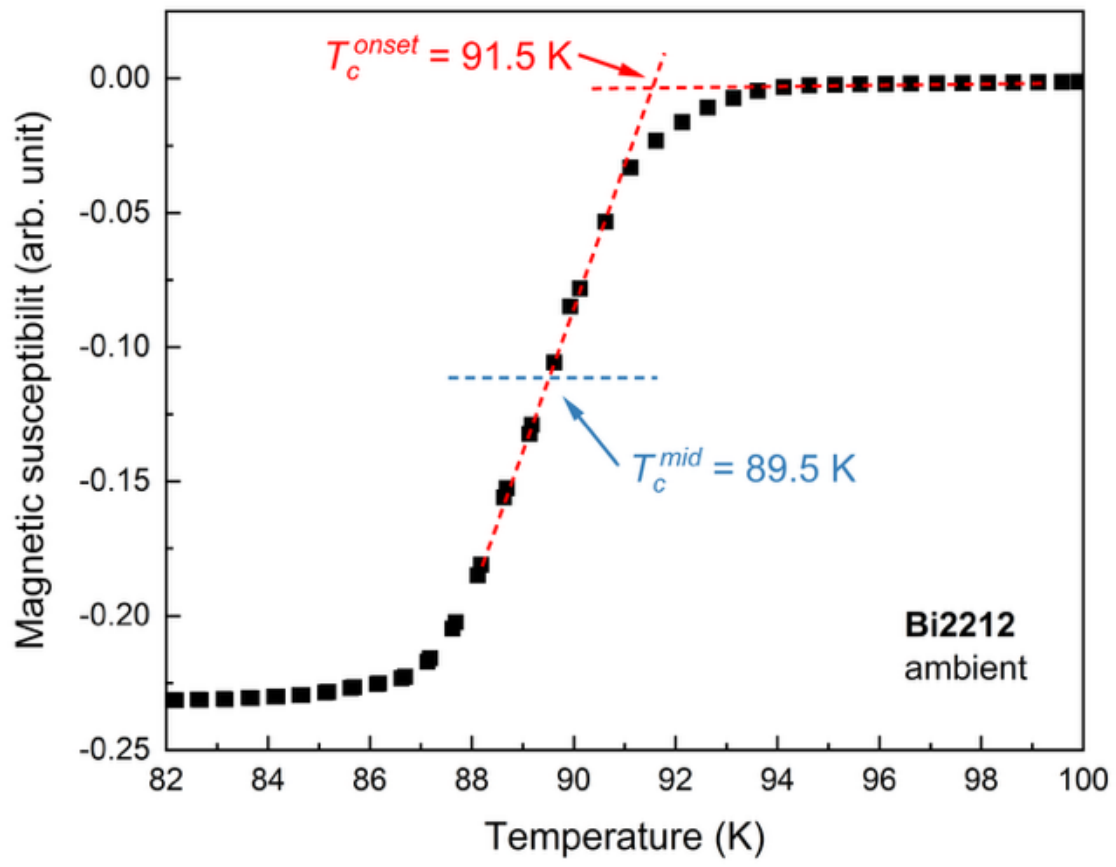

**Extended Data Fig. 1 | Magnetic characterization of bulk superconductivity.** Temperature dependence of the AC magnetic susceptibility for the Bi2212 single crystals. The sharp diamagnetic transition, featuring an onset temperature $T_c^{onset} \approx 91.5$ K and a midpoint $T_c^{mid} \approx 89.5$ K, verifies the bulk superconductivity of the grown sample batch. Measurements were performed using a Quantum Design Magnetic Property Measurement System (MPMS) with an AC field amplitude of 1 Oe at a frequency of 100 Hz.

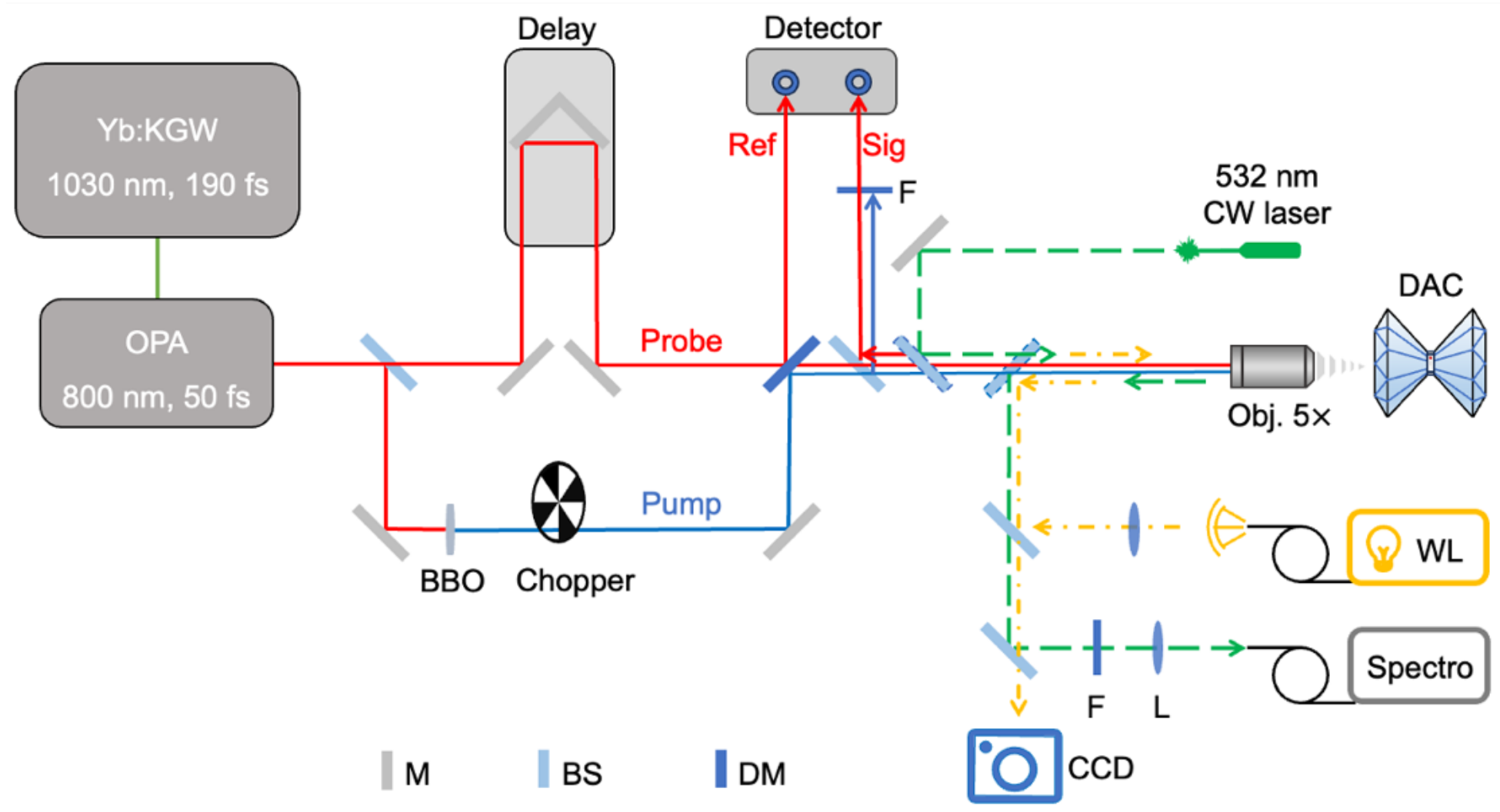

**Extended Data Fig.2 | Optical setup for high-pressure ultrafast spectroscopy.** A single long-working-distance objective lens (5×, $NA = 0.14$) focuses the incident light and collects the reflected signals through the upper diamond anvil in a collinear geometry. Flip mirrors within a cage system (dashed outlines) selectively direct the optical path into three functional modules: (i) two-color pump-probe spectroscopy, where a dichroic mirror coaxially combines the 400 nm pump and 800 nm probe pulses; (ii) sample imaging under white-light illumination; and (iii) *in situ* pressure calibration via 532 nm continuous-wave excitation and ruby photoluminescence (PL) collection. OPA, optical parametric amplifier; BBO, $\beta$-barium borate; M, mirror; BS, beam splitter; DM, dichroic mirror; CCD, charge-coupled device; F, filter; L, lens; DAC, diamond anvil cell; WL, white light; Obj., objective lens.

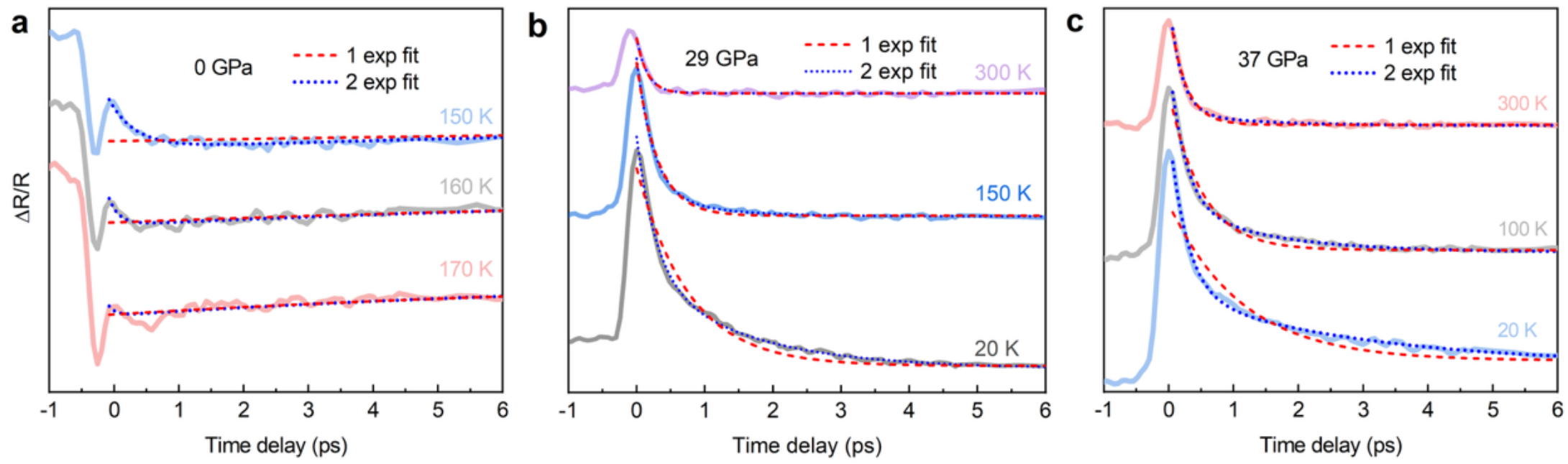

**Extended Data Fig. 3 | Validation of the multi-component decay analysis. a,** Representative transient reflectivity traces at ambient pressure near the pseudogap temperature. **b, c,** Corresponding analysis at 29 GPa (b) and 37 GPa (c) at representative temperatures. In all panels, experimental data (solid lines) are compared with single-exponential (dashed lines) and double-exponential (dotted lines) fits. A bi-exponential model is required to capture the sharp initial recovery near zero delay, providing a consistent description of the coexisting two relaxation channels across all measured pressures.

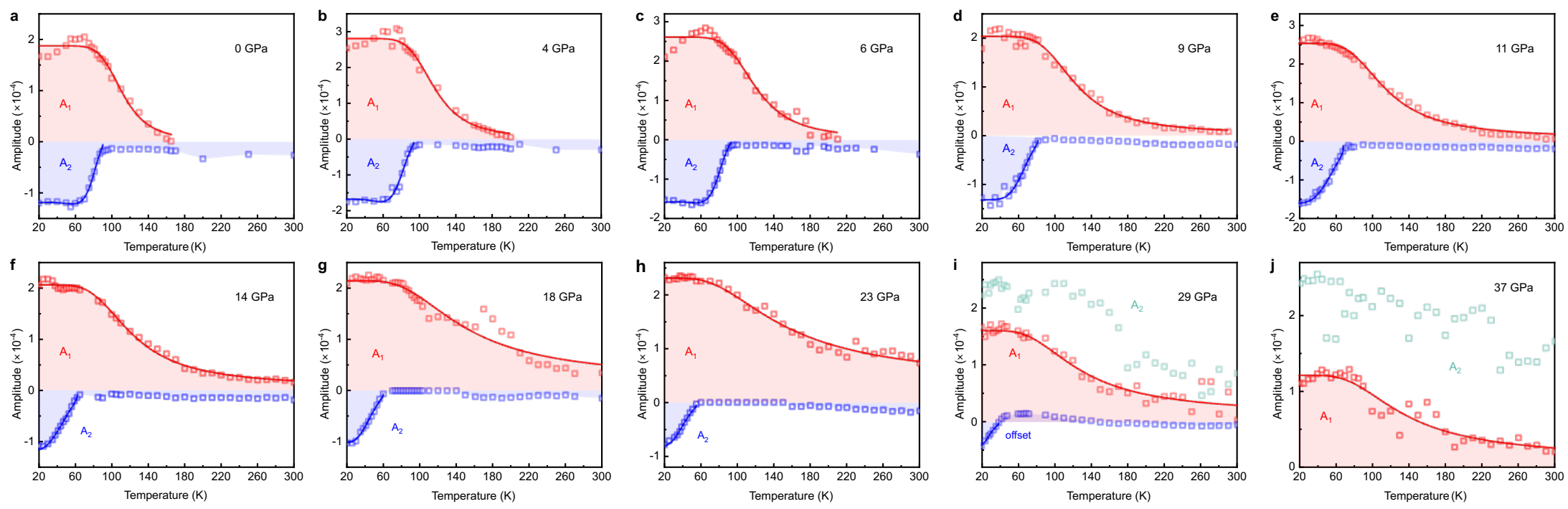

**Extended Data Fig. 4 | Complete pressure evolution of quasiparticle amplitudes. a-j,** Temperature dependence of the pseudogap amplitude $A_1$ and superconducting amplitude $A_2$, extracted using the unified bi-exponential decay model across all ten investigated pressures. Solid curves represent the fits to the phenomenological Rothwarf-Taylor model (Eqs. 1 and 2). Data at 0, 6, 14, 23, 29, and 37 GPa (replotted from Fig. 3) are included to delineate the continuous phase evolution. The amplitude profiles reveal the gradual suppression of the superconducting component before its complete disappearance and the subsequent emergence of a positive component at 37 GPa.

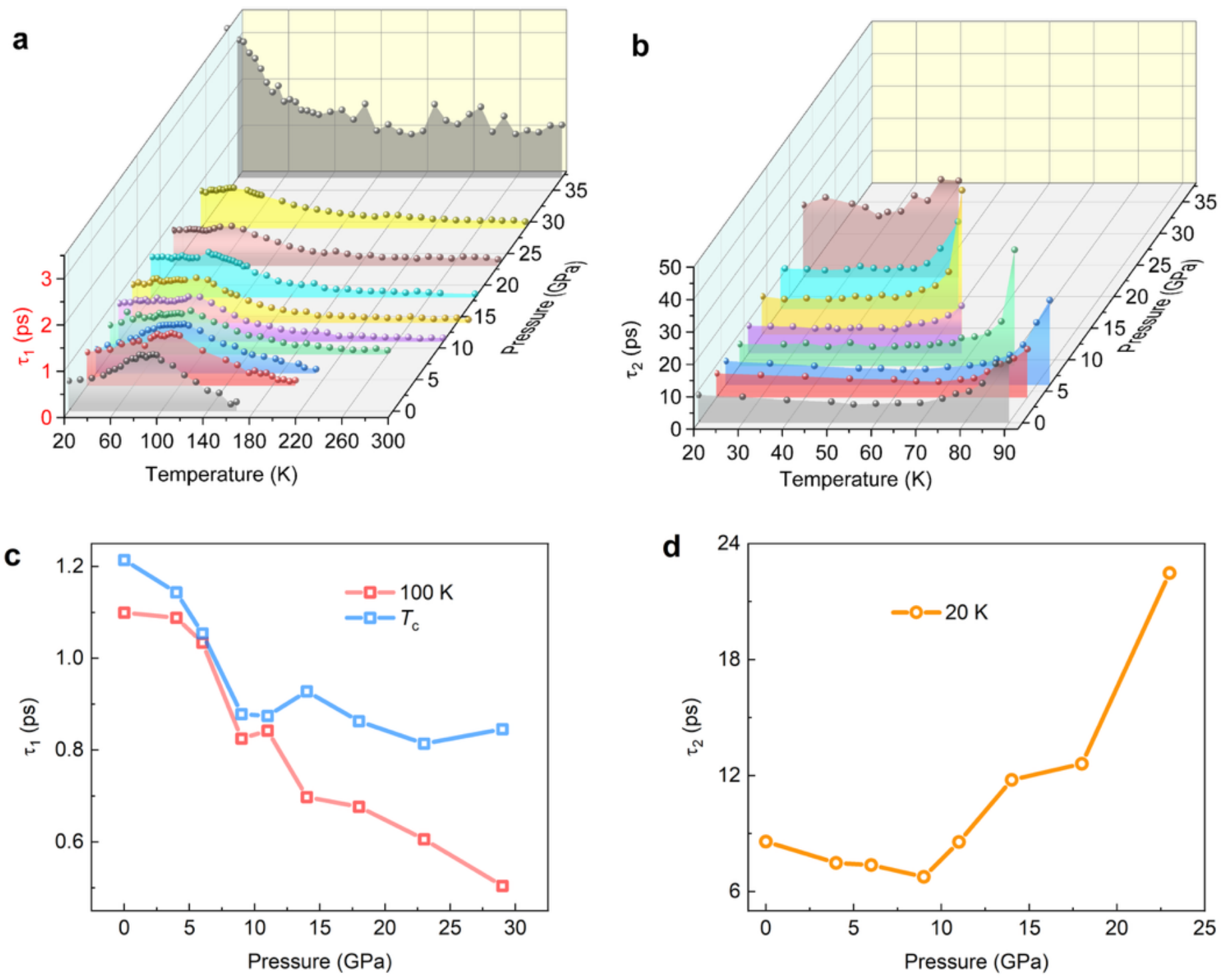

**Extended Data Fig. 5 | Pressure evolution of quasiparticle recombination lifetimes.** **a, b,** Waterfall plots of the temperature and pressure dependence of the relaxation times for the pseudogap $\tau_1$ (**a**) and superconducting $\tau_2$ (**b**) components. Panel **b** isolates the response below $T_c$ to highlight the divergence of the slow recombination lifetimes. **c,** Pressure dependence of $\tau_1$ extracted at 100 K and at $T_c$. The abrupt drop in lifetime between 6 and 10 GPa arises from the expansion of the available scattering phase space, driven by the suppression of the pseudogap $\Delta_{PG}$. **d**, Pressure dependence of $\tau_2$ at 20 K. The enhancement of $\tau_2$ under compression is attributed to the reduction of the superconducting gap, consistent with the phonon bottleneck mechanism in the RT model[34,35].

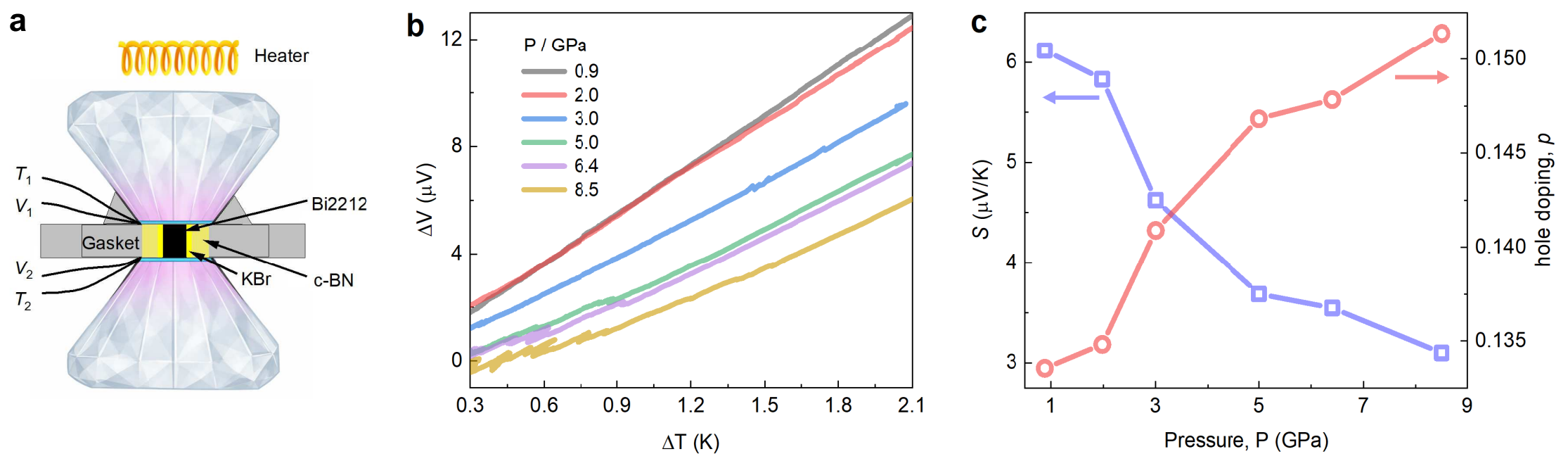

**Extended Data Fig. 6 | High-pressure thermoelectric characterization. a,** Schematic of the experimental configuration. A resistive micro-heater establishes a thermal gradient across the sample, with local temperatures ($T_1$, $T_2$) monitored by thermocouples, while the corresponding voltages ($V_1$, $V_2$) are collected via platinum leads. Cubic boron nitride (c-BN) powder provides electrical insulation from the metal gasket. **b,** Thermoelectric voltage measurements. Representative raw data displaying $\Delta V = V_1 - V_2$ as a function of $\Delta T = T_1 - T_2$. The slope of the linear response yields the uncorrected Seebeck coefficient ($S_{meas}$). **c,** Pressure evolution of the room-temperature Seebeck coefficient ($S$, left axis) and the derived hole concentration ($p$, right axis). The intrinsic Seebeck coefficient $S$ is obtained by subtracting the platinum lead contribution ($S_{Pt} \approx -5\ \mu V/K$)[57] via $S = S_{meas} - S_{Pt}$. The hole concentration is calculated using the empirical relation $S = 992\exp(-38.1p)$[58]. Measurements were restricted to pressures below 9 GPa due to pressure-induced sample fracture.